\documentclass[12pt]{iopart}
\usepackage{iopams}
\usepackage{setstack}
\usepackage{graphicx}
\begin{document}
\title[Laser-induced solid-solid phase transition in As under pressure]
{Laser-induced solid-solid phase transition in As under pressure: 
A theoretical prediction}
\author{Eeuwe S.\ Zijlstra$^1$, Nils Huntemann$^1$ and Martin E.\ Garcia$^1$}
\address{$^1$ Theoretische Physik, Universit\"at Kassel, Heinrich-Plett-Str.\ 40, 34132 Kassel, Germany}
\ead{Zijlstra@physik.uni-kassel.de}
\begin{abstract}
In Arsenic a pressure-induced solid-solid phase transition from the 
A7 into the simple cubic structure has been experimentally demonstrated 
[Beister et al. 1990 \textit{Phys.\ Rev.\ B} \textbf{41} 5535].
In this paper we present calculations, which predict
that this phase transition can also be induced by an ultrashort laser pulse in As under pressure.
In addition, calculations for the pressure-induced phase transition are presented.
Using density functional theory in the generalized gradient approximation, 
we found that the pressure-induced phase transition takes place at $26.3$ GPa and is accompanied
by a volume change $\Delta V = 0.5$ $a_0^3$/atom.
The laser-induced phase transition is predicted for an applied pressure of $23.8$ GPa
and an absorbed laser energy of $2.8$ mRy/atom. 
\end{abstract}
%
%
\submitto{\NJP}
\maketitle

\section{Introduction}

Femtosecond lasers create extreme nonequilibrium conditions in matter, namely, 
one with hot electrons and cold ions \cite{VanVechten79}.
This is primarily because laser light interacts very strongly with electrons but not with ions, and
secondarily because the electrons thermalize relatively slowly with the ions 
compared with the time needed for ultrafast structural changes (typically several ps vs several $100$'s of fs).
The possibility of heating the electrons in a material with an ultrafast laser pulse without immediately
changing the temperature of the ions has paved the way to the discovery of interesting physical phenomena,
such as, perhaps most importantly, laser-induced phase transitions that cannot be explained by the
heating of the ions, but are caused by a change of the electronic bonding properties.

In Arsenic a series of pressure-induced phase transitions exists.
At ambient conditions As crystallizes in the A7 structure. 
At $25$ GPa there is a transition to the simple cubic (sc) phase \cite{Beister90},
followed by transitions \cite{Greene95} to the so-called As(III) structure at $48$ GPa and
to body-centered cubic As at $97$ GPa. 
The fact that As in the A7 structure is electronically stabilized by the Peierls mechanism \cite{Shang07}
makes it a candidate for an ultrafast laser-induced solid-solid phase transition.
This is of fundamental interest, because the majority of the experimentally studied laser-induced phase transitions
shows ultrafast melting, for example, in Si \cite{Shank83}, GaAs \cite{Siegal95},
Te \cite{Ashitkov02}, and InSb \cite{Lindenberg05}.
Examples of laser-induced solid-solid phase transitions are 
a disorder-to-disorder transition in amorphous GeSb \cite{Callan01}, 
a metallic-to-semiconductor phase transition in SmS \cite{Kitagawa03},
a monoclinic-to-rutile transition in VO$_2$ \cite{Cavalleri01,Kuebler07},
a ferromagnetic-to-paramagnetic transition in CoPt$_3$ alloy films \cite{Beaurepaire98},
and an antiferromagnetic-to-ferromagnetic phase transition in FeRh \cite{Ju04}.
The main topic of this paper
is to analyze the possibility to induce a solid-solid phase transition by an ultrafast laser pulse
in As under pressure.

\begin{figure}
  \begin{indented}
  \item[]\includegraphics[width=12cm]{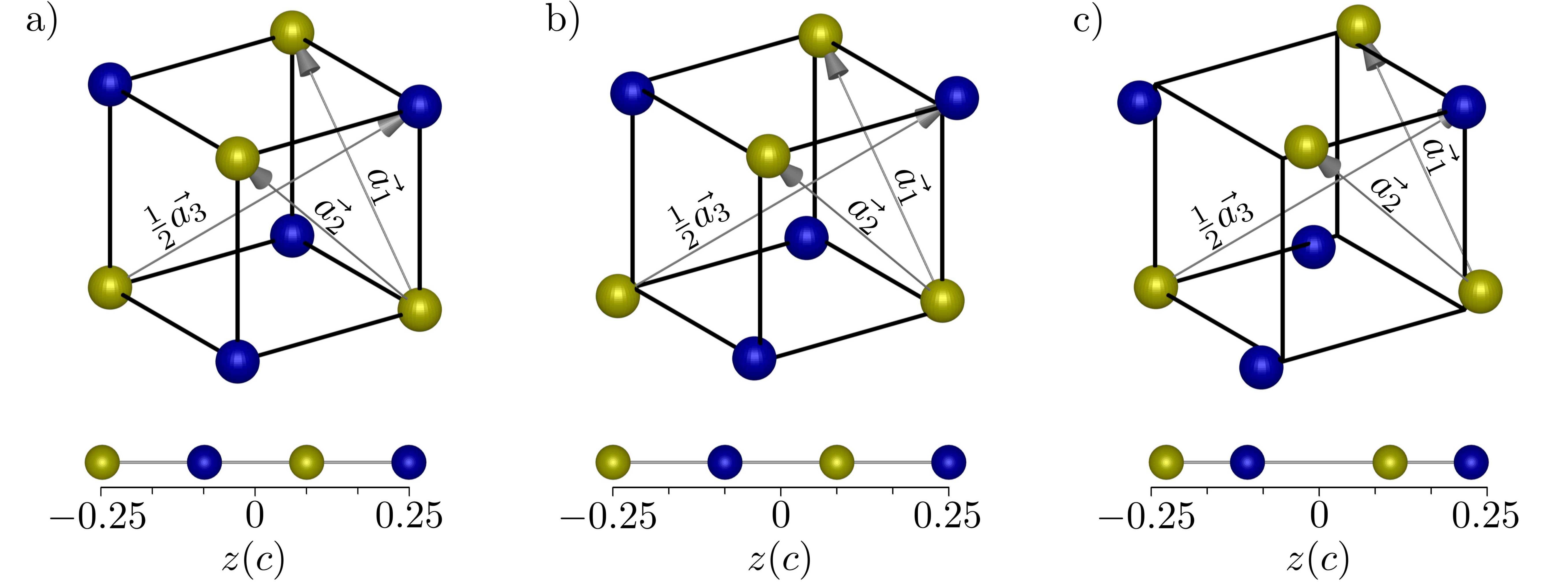}
  \end{indented}
  \caption{\label{fig_A7}
  Relation between the A7 and sc structures.
  a) The sc structure. 
  The lattice vectors $\mathbf{a}_1$, $\mathbf{a}_2$, and $\mathbf{a}_3$ 
  belong to the A7 structure (the sc structure is a special
  case of the A7 structure).
  b) Intermediate structure.
  Compared to a)
  the unit cell of the sc lattice is elongated along the vector $\mathbf{a}_3$ keeping
  the volume per atom constant.
  The new positions of the atoms are shown together with a cube of the sc lattice.
  c) The A7 structure of As at ambient pressure.
  Compared to b)
  the atomic planes perpendicular to $\mathbf{a}_3$ are displaced alternatingly
  in the $\mathbf{a}_3$ and $-\mathbf{a}_3$ directions due to a Peierls distortion.
  The cube still represents the sc structure of a), for reference.
  Below a) -- c) projections of the atomic planes onto $\mathbf{a}_3$ are shown.}
\end{figure}

We now describe the relationship between the A7 and sc structures, the two phases of As between which we  
will predict a laser-induced transition (see section \ref{sec_results}).
Starting from an sc atomic packing, the A7 structure can be derived in two steps, 
which are illustrated in \fref{fig_A7}.
First the sc lattice is deformed by elongating it along one of the
body diagonals (indicated by $\frac{1}{2} \mathbf{a}_3$ in figure \ref{fig_A7}).
The magnitude of this deformation is usually expressed by $c/a$, where $c$ is the length
of the lattice vector $\mathbf{a}_3$ and $a$ is the length of $\mathbf{a}_1$ (\fref{fig_A7}).
In the sc lattice $c/a = \sqrt 6 \approx 2.45$.
In the second step a Peierls instability causes the atoms to be displaced along the
same diagonal, in opposite directions (\fref{fig_A7}).
The magnitude of the displacements is expressed by the atomic coordinate $z$:
A value of $0.25$ indicates no Peierls distortion. Deviations from $z = 0.25$
give the magnitude of the displacement of the atoms in the $\mathbf{a}_3$ direction.
In As at ambient pressure $z = 0.228$ \cite{Schiferl69},
which leads to a displacement of the atoms  
amounting to $13\%$ of the average distance between the planes [see \fref{fig_A7}(c)].

\section{\label{sec_method}Method}

To determine the phase of As under pressure before the laser excitation
when the electrons are in their ground state we compared the enthalpies of the A7 and sc phases, which were computed
in the following way.
(i) We calculated total energies for a series of unit cell 
volumes with the computer program \textsc{wien2k} \cite{wien2k_7.1a}.
This is an all-electron full-potential linearized augmented plane wave code, which has been designed
to make very accurate predictions for solids within the limitations of density functional theory \cite{Kohn65}.
For the sc structure these calculations are straightforward, because there are no internal parameters.
However, for the A7 structure the energy is a function of both volume and the internal parameters
$c/a$ and $z$ described in the Introduction.
In this case
we minimized the total energy at each volume by
optimizing the $c/a$ ratio and the Peierls distortion parameter $z$.
(ii) We fitted the total energy vs volume data for the sc phase to the
Birch-Murnaghan equation of state \cite{Birch47}, which has been derived for systems with cubic symmetry.
The fitting parameters of this equation of state are the minimal total energy $E_0$,
the corresponding atomic volume $V_0$,
the bulk modulus $B_0$, and the first derivative of the bulk modulus with respect to
pressure $B_0'$.
(iii) We fitted the computed energy differences between the A7 and sc structures to
\begin{equation}
  \label{eq_power}
  E_{A7} - E_{sc} =\cases{0                                             &for $V < V_c$\\
                          A_c \left( \case{V}{V_c} - 1 \right)^{\beta_c}&for $V > V_c$.\\}
\end{equation}
We found that this form described our A7 data much better than when
the total energies of the A7 structure are fitted directly to the Birch-Murnaghan equation of state \cite{Birch47}.
(iv) We calculated the enthalpies $H = E + P V$ for the sc and A7 structures
from the analytical fits for the total energy 
and $P = - \left( \frac{\partial E}{\partial V} \right)_S$.

To describe As after laser excitation we use the following physical picture: 
The laser pulse creates electrons and holes, which
undergo dephasing and collisions on a timescale that is much shorter than the typical time of ionic motion
($\sim 130$ fs, based on the $A_{1g}$ optical $\Gamma$-point phonon frequency of As in its equilibrium structure).
Therefore one can for all practical purposes assume that the excited carriers thermalize instantaneously.
The effect of the excitation by an ultrashort laser pulse was thus simulated by heating the electrons.
In this case, the phase stability is governed by the free energy $F = E - T S$, where
$T$ is the electronic temperature and $S$ is the electronic entropy
(To be sure, the free energy was also used for the ground state calculations,
but since the electronic temperature
was very low in that case and was used for smearing purposes only, we have ignored this point for the sake
of clarity of presentation in the discussion above).
Note that the entropic contribution of the ions to the free energy was neglected.
This is because we assumed that both the laser
pulse duration and the timescale on which the phase transition takes place are much shorter than the 
electron-ion interaction time,
so that no substantial heating of the ions due to the laser occurs during the time of interest.
Another consequence of the short timescales that we are looking at is that the system does not have time to
expand, so that structural changes should be assumed to take place at constant volume.
Therefore, we compared the free energies of the A7 and sc structures
to predict under which conditions a phase transition takes place in As under pressure.

Details of our \textsc{wien2k} calculations were as follows.
We used the generalized gradient approximation \cite{Perdew96}
for the exchange and correlation energy.
Our basis consisted of plane waves with energies less than or equal to
$20.25$ Ry.
Inside the so-called muffin-tin spheres around each atom the plane waves were
augmented by atomic orbitals with a linearized energy dependence.
To get a better convergence with respect to the number of plane waves included in the basis
the \textit{3d}, \textit{4s}, and \textit{4p} states were described 
with a combination \cite{Madsen01} of energy independent augmented plane waves and local
orbitals.
Also, additional local orbitals \cite{Singh91} were added for the \textit{3d} and the \textit{4s} states.
The first Brillouin Zone of the A7 structure was sampled using a grid of $32 \times 32 \times 32$ k points
excluding the $\Gamma$ point, which corresponds to $2992$ irreducible k points.
To rule out errors due to, for example, a different k-space sampling and to
allow for a detailed comparison between the A7 and the sc structures, the
latter structure was also computed as an A7 structure with the parameters $c / a$ and $z$ fixed
to their sc values $\sqrt 6$ and $0.25$, respectively.

The electronic ground state calculations were performed using temperature smearing
($T = 1$ mRy) to determine the electronic occupation numbers.
For the laser-excited state an electronic temperature of $T = 19$ mRy was used.
This choice corresponds to an absorbed laser energy of $E_{laser} \approx 2.8$ mRy/atom or,
equivalently, the creation of $\approx 0.06$ electron-hole pairs per unit cell (when the atomic
volume $V = 114.0$ $a_0^3$/atom).
Since we assumed an instantaneous thermalization of the excited carriers, our predicted results
do not depend directly on the energy of the phonons, but only on the total absorbed energy per atom.
The incident laser fluence that leads, at the sample surface, to the above-mentioned value of $E_{laser}$
is, however, a function of both the reflectivity and the penetration depth of the laser light, which
are dependent on the laser wavelength.
A complication is that these dependencies might change due to an applied pressure and under the influence of the laser
irradiation.
Ignoring these latter points and using a laser wavelength of $800$ nm we estimated our incident fluence 
to be $\sim 1.8$ mJ/cm$^2$ (for this wavelength the reflectivity is $\sim 55\%$ \cite{Cardona64} and the 
penetration depth is $\sim 23$ nm, based on the experimentally determined complex dielectric constant 
of $16 + 27i$ \cite{Raisin74}).
Here we would further like to mention that we assumed the laser pulse duration to be much shorter than the
timescale of ionic motion ($\sim 130$ fs).

\section{\label{sec_results}Results}

Before turning to the energy vs volume curves, a few words about the equilibrium structure of Arsenic 
are at place, because this allows us to get a rough idea about the quality of the predictions
that the GGA \cite{Perdew96} we used is able to make.
According to our calculations the ground state parameters of As in the A7 structure
are $c/a = 2.86$, $z = 0.226$, and $V = 154$ $a_0^3$/atom.
This is in reasonable agreement with the measured values
$c/a = 2.78$, $z = 0.228$, and $V = 144$ $a_0^3$/atom \cite{Schiferl69}.
As was to be expected,
our atomic volume $V$ is larger than the experimental value \cite{Schiferl69} 
by about $7\%$, which
is consistent with the general trend of the GGA to overestimate bond lengths, and
in agreement with the value of $154$ $a_0^3$/atom recently obtained from GGA calculations 
by Shang and co-workers \cite{Shang07}. 
From calculations for smaller atomic volumes we concluded that
our too large values for the magnitude of the Peierls distortion ($0.25 - z$) and for $c/a$
can in part but not entirely be understood as a consequence of the overestimated unit cell volume.

\begin{figure}
  \begin{indented}
  \item[]\includegraphics[angle=90,width=8cm]{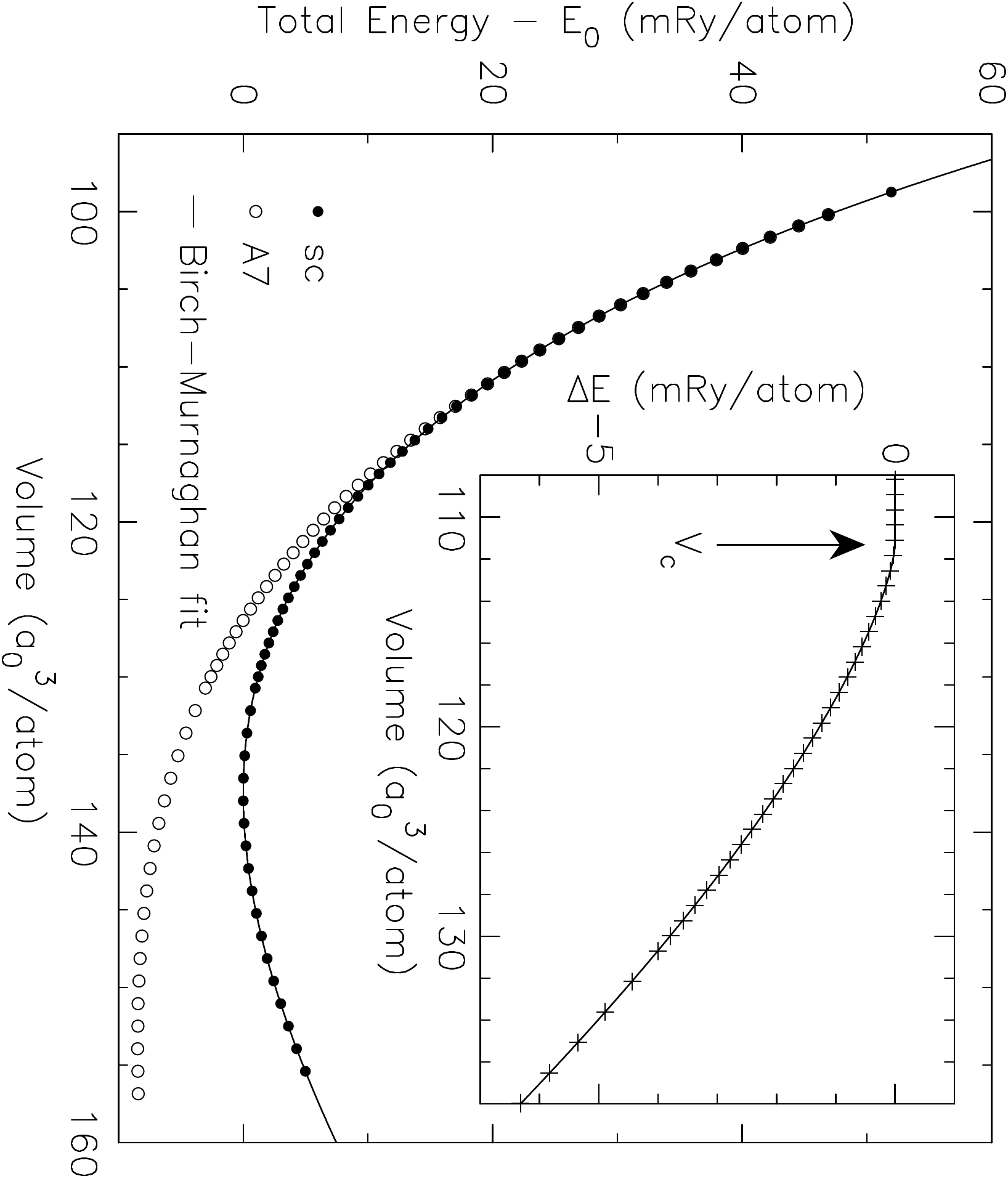}
  \end{indented}
  \caption{\label{fig_fit}
  Total ground state energy vs volume for sc As and As in the A7 structure.
  The solid curve shows the Birch-Murnaghan fit to the data for the sc structure.
  The inset shows the energy difference $\Delta E = E_{A7} - E_{sc}$ (plusses) and the 
  fit to \eref{eq_power} (solid curve).
  The volume $V_c$ at which the A7 structure becomes energetically favored over the
  sc one is indicated.}
\end{figure}

The total energies of the sc and A7 structures as a function of atomic volume
are shown in figure \ref{fig_fit}.
These data are for the electronic ground state, i.e., 
before laser excitation.
The parameters obtained from the fits to the Birch-Murnaghan equation of state
and \eref{eq_power}, which we made using the data points plotted in figure \ref{fig_fit}
and the inset of figure \ref{fig_fit}, respectively, are given in table \ref{table_fit}.
\begin{table}
  \caption{\label{table_fit}
  Fitting parameters for the electronic ground state ($T = 1$ mRy) and the 
  laser-excited state ($T = 19$ mRy). 
  $V_0$, $B_0$, and $B_0'$ give the best fit of our sc data
  to the Birch-Murnaghan equation of state \cite{Birch47}.
  $V_c$, $\beta_c$, and $A_c$ represent our best fit of the computed energy differences 
  between the A7 and the sc structures to \eref{eq_power}.} 
  \begin{indented}
  \item[]\begin{tabular}{ccc}
  \br
                       & $T = 1$ mRy & $T = 19$ mRy \\
  \mr
  $V_0$ ($a_0^3$/atom) & $137.479$   & $138.265$    \\
  $B_0$ (GPa)          & $78.107$    & $76.241$     \\
  $B_0'$               & $4.317$     & $4.341$      \\
  $V_c$ ($a_0^3$/atom) & $111.31$    & $115.54$     \\
  $\beta_c$            & $1.429$     & $1.611$      \\
  $A_c$ (mRy/atom)     & $-48.63$    & $-59.37$     \\
  \br
  \end{tabular}
  \end{indented}
\end{table}
From figure \ref{fig_fit} it is apparent that our fits follow the computed data 
very closely.
Indeed, the root mean square of the residuals was only $0.003$ mRy/atom
for the Birch-Murnaghan fit and $0.005$ mRy/atom for the fit to \eref{eq_power}.

\begin{figure}
  \begin{indented}
  \item[]\includegraphics[angle=90,width=8cm]{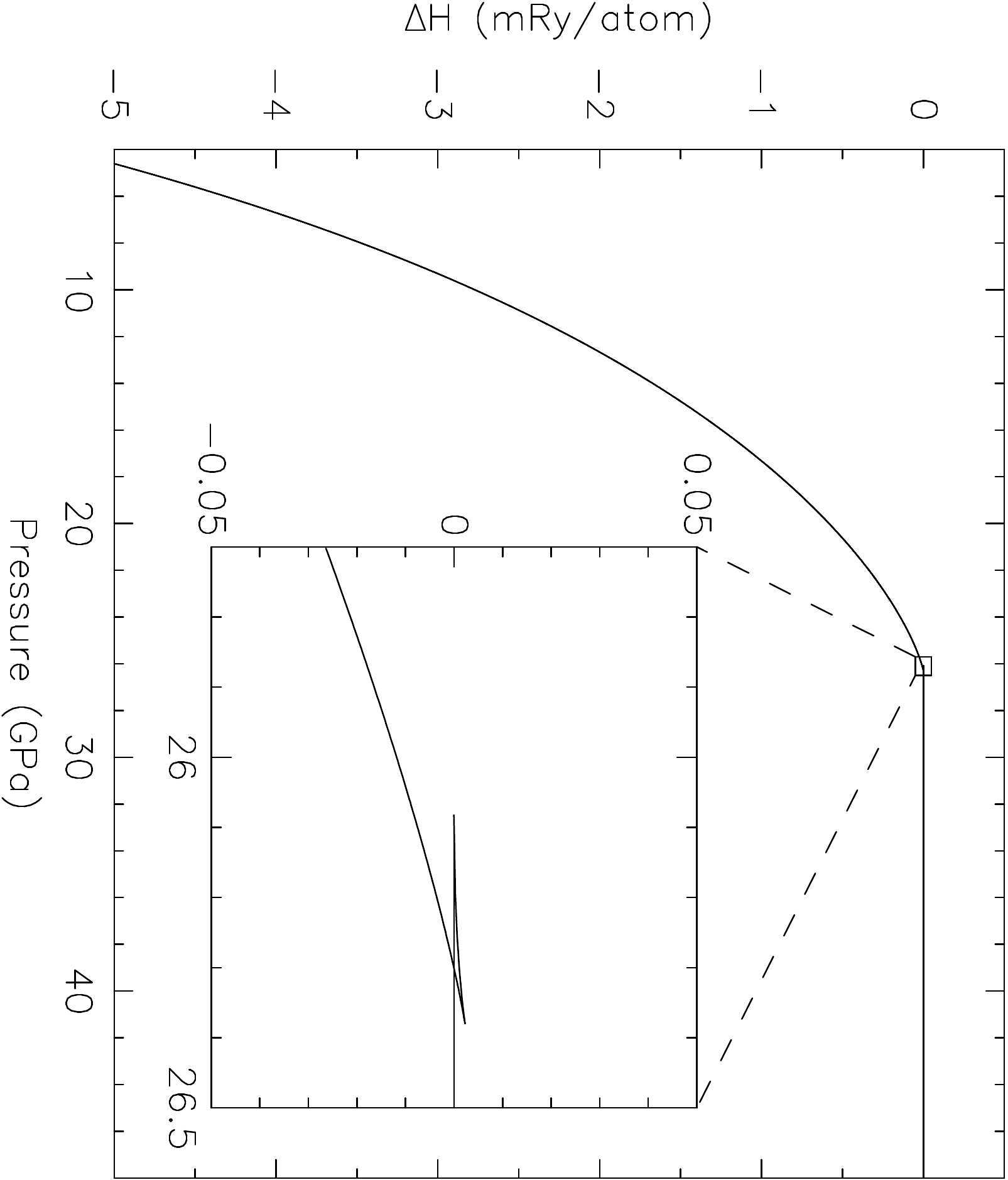}
  \end{indented}
  \caption{\label{fig_enth}
  Enthalpy difference $\Delta H = H_{A7} - H_{sc}$ as a function of pressure.
  The inset shows the indicated region enlarged.
  The transition from the A7 to the sc structure takes place at a pressure of $26.3$ GPa.}
\end{figure}

The enthalpy of the A7 structure relative to the
enthalpy of the sc structure as a function of pressure is plotted in 
figure \ref{fig_enth}.
In a narrow pressure range multiple solutions exist for the enthalpy of the A7 structure, as
is shown in the inset of figure \ref{fig_enth}.
Of course, the solution with the lowest enthalpy is the stable one.
A consequence of these multiple solutions is 
that the enthalpies of the A7 and sc structures cross and that the transition point
from the A7 to the sc structure can thus easily be determined.
From figure \ref{fig_enth} we see that the transition takes place at $26.3$ GPa,
in reasonable agreement with earlier GGA calculations by H\"aussermann and co-workers \cite{Haussermann02},
who found the transition pressure to be $28$ GPa.
Another consequence of the multiple solutions seen in the inset of figure \ref{fig_enth} is
that the unstable solutions, corresponding to 
a narrow range of possible atomic volumes, are not realized for any given pressure,
which means that there is a small volume change at the transition, which according to our
calculations is $\Delta V / V_c = 0.4\%$.
Both the transition pressure and the volume change are in nearly perfect agreement with
the experimental values of Beister and co-workers \cite{Beister90}, who have found
the transition at a pressure of $25 \pm 1$ GPa accompanied by a volume 
change $\Delta V / V_c \lesssim 0.6\%$.
We did not find a discontinuity in $c/a$ at the transition,
which according to Beister and co-workers \cite{Beister90} jumps from $2.49$ to
$\sqrt 6 \approx 2.45$.
However, according to our calculations $z$ changes discontinuously from $0.246$ to $0.25$.

\begin{figure}
  \begin{indented}
  \item[]\includegraphics[angle=90,width=8cm]{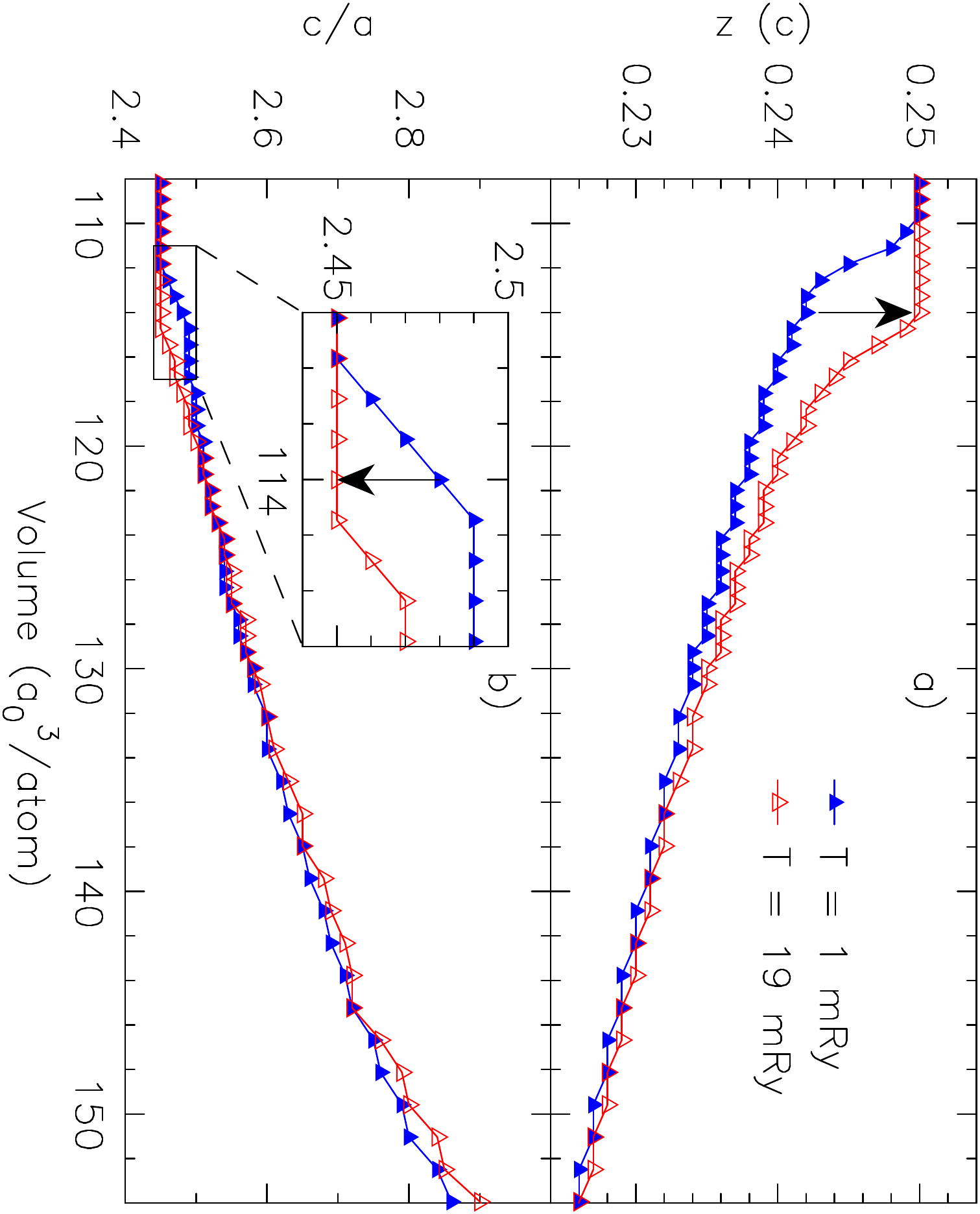}
  \end{indented}
  \caption{\label{fig_params}
  a) the Peierls parameter $z$ and b) the ratio $c/a$
  as a function of volume before ($T = 1$ mRy)
  and after laser excitation ($T = 19$ mRy).
  The inset of b) shows a selected region enlarged.
  Arrows in a) and the inset of b) indicate the possibility of a laser-induced phase transition at constant volume
  from the A7 to the sc structure.}
\end{figure}

\begin{figure}
  \begin{indented}
  \item[]\includegraphics[width=8cm]{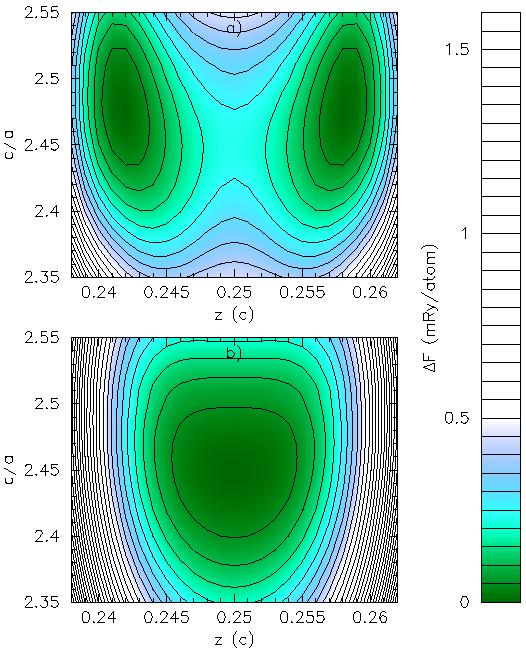}
  \end{indented}
  \caption{\label{fig_pes}
  Potential energy surface contour plot of As a) before and b) after laser-excitation
  as a function of the internal parameters $z$ and $c/a$ at
  the volume $V = 114.0$ $a_0^3/$atom.
  Whereas
  the energy minimum of As before laser-excitation is relatively sharp (the two minima shown are equivalent),
  the minimum after laser-excitation is broader. 
  The minimum in b) is located at the sc values of $z$ and $c/a$ ($0.25$ and 
  $\sqrt 6 \approx 2.45$, respectively).}
\end{figure}

We now describe the effect of an ultrashort laser pulse.
As we have already argued in section \ref{sec_method} the timescale that we are interested in
is too short for the system to expand. 
Therefore we present results at constant volume, and
we only briefly discuss the effect of pressure at the end of this paragraph.
Our optimized $c/a$ ratio and Peierls parameter $z$ as a function of volume for As in the A7 structure
are shown in figure \ref{fig_params}.
It can be seen that the pressure-induced phase transition from the A7 to the sc structure occurs at a larger volume
when the electrons have been heated with a laser compared to when they are in their ground state.
As a consequence, it is possible to induce a phase transition at constant volume from the A7 to the sc structure
in As under pressure.
We have indicated this possibility for a specific volume ($V = 114.0$ $a_0^3$/atom)
by two arrows in figure \ref{fig_params}.
The corresponding potential energy surfaces before and after laser excitation are shown in \fref{fig_pes}.
When the electrons are in the ground state this volume corresponds to an applied calculated pressure
of $23.8$ GPa. 
The optimized ratio $c/a = 2.48$ and the Peierls parameter $z=0.242$ indicate that the system is in the A7 phase.
After the laser excitation the optimal $c/a$ and $z$ become $2.45$ and $0.25$, respectively,
signifying that As has made the transition to the sc structure.
At this given volume, the internal pressure in laser-excited As has dropped to $22.4$ GPa.
If the system were given enough time to adjust to the applied pressure of $23.8$ GPa, the volume 
would, according to our calculations, decrease to $113.1$ $a_0^3$/atom, in which case the simple cubic phase would clearly
still be the most stable one (see figure \ref{fig_params}).

Comparing the free energies of the A7 and sc structures it follows that the A7 to sc transition in As can,
in principle, be induced for atomic volumes up to $V \leq V_c = 115.54$ $a_0^3/$atom 
($T = 19$ mRy data in table \ref{table_fit}), which corresponds to an applied pressure of $\geq 22.1$ GPa.
However, near the transition point the potential energy surface of As becomes very flat (see \fref{fig_pes}),
and as a consequence the uncertainty in the optimized $c/a$ and $z$ values becomes relatively large,
which explains why in figure \ref{fig_params} the $c/a$ ratio and $z$ Peierls parameter start
to deviate from their sc values at volumes less than $V_c = 115.54$ $a_0^3/$atom
(but greater than $V = 114.0$ $a_0^3/$atom).
So, in order to be on the safe side and to avoid any ambiguity about the final state after
the laser excitation, the lowest pressure for which we feel confident to predict an unambiguous
transition to the sc structure (as defined by its $c/a$ and $z$ values) is the value of $23.8$ GPa
derived in the previous paragraph.

As can be noted in figure \ref{fig_params}(b), the predicted phase transition involves a change in $c/a$.
Since the volume remains constant, in a polycrystalline sample there is no average effect and we expect
that the change in $c/a$ takes place on a subpicosecond timescale.
In a monocrystal this change might be limited by the sound velocity.

\section{Conclusion}

Summarizing, we demonstrated that it is possible to induce a solid-solid phase transition in As
under pressure using an ultrashort laser pulse.
In particular, for an absorbed laser energy $E_{laser} \approx 2.8$ mRy/atom,
we predicted that the phase transition will be induced under an applied pressure as low as $\geq 23.8$ GPa.
For higher absorbed energies we expect that the transition could be induced under an even lower
applied pressure.
This transition might be observed using time-resolved x-ray diffraction experiments.

\ack

This work has been supported by the Deutsche Forschungsgemeinschaft (DFG)
through the priority program SPP 1134 and by the European Community Research
Training Network FLASH (MRTN-CT-2003- 503641).

\section*{References}


\end{document}